\documentclass[letter,onecolumn]{revtex4}

\usepackage{amsmath}
\usepackage[dvips]{graphicx}
\usepackage{amssymb}

\begin{document}
\title{3D Dune Skeleton Model as a Coupled Dynamical System of 2D Cross-Sections}

\author{Hirofumi Niiya}
\affiliation{Department of Mathematical and Life Sciences,  Hiroshima University, Higashi-Hiroshima 739-8526}
\author{Akinori Awazu}
\affiliation{Department of Mathematical and Life Sciences,  Hiroshima University, Higashi-Hiroshima 739-8526}
\author{Hiraku Nishimori}
\affiliation{Department of Mathematical and Life Sciences,  Hiroshima University, Higashi-Hiroshima 739-8526}

\begin{abstract}
To analyze theoretically the stability of the shape and the migration process of transverse dunes and barchans, we propose a {\it skeleton model} of 3D dunes described with coupled dynamics of 2D cross-sections.
First,
2D cross-sections of a 3D dune parallel to the wind direction are extracted as elements of a skeleton of the 3D dune,
hence, the dynamics of each and interaction between them is considered.
This model simply describes the essential dynamics of 3D dunes as a system of coupled ordinary differential equations.
Using the model we study the stability of the shape of 3D transversal dunes and their deformation to barchans depending on the amount of available sand in the dune field, sand flow in parallel and perpendicular to wind direction.
\end{abstract}


\maketitle

Sand dunes, which are the largest granular objects on the Earth, move by wind and exhibit various morphodynamics.
As typical shapes of dunes, barchan, transverse dune, linear dune, star dune, dome dune and parabolic dune are known\cite{mckee1979introduction,cooke1993desert}.
The steadiness of wind direction and the amount of available sand in each dune field are considered as dominant  factors determining these shapes.
For example, unidirectional steady wind generates barchans or transverse dunes.
The former are crescentic shaped dunes, and are formed in dune fields with small amounts of available sand, whereas transverse dunes which extend perpendicular to the wind direction, are formed in dune fields with the larger amount of available sand than the barchan-rich field.
A characteristic aspect of recent dune studies is that quantitative analysis of dune morphodynamics has largely progressed.
In particular, water tank experiments and computer models have uncovered the complex processes of dunes\cite{hersen2002relevant,bishop2002modelling,endo2005barchan,katsuki2005emergence,katsuki2005collision,groh2008barchan}.
They were successful to reproduce formation and migration processes of barchans and other types of dunes under controlled setups.
However, theoretical methodology to explain the complex morphodynamics of dunes beyond only reproducing them is yet to be developed.
We, here, propose a {\it skeleton model} of 3D dunes described with coupled dynamics of 2D cross-sections\cite{nishimori2009abcde},
which has a form of coupled ordinary differential equations.
Using the model we study the morphodymamics of dunes,
particularly the stability of the shape of transverse dunes and its deformation to barchans.

Recently Katsuki and Nishimori has proposed a model for the collision dynamics of two 3D barchans focusing on the dynamics of their central 2D cross sections [8][9].
That is called ABCDE(Aeolian/Aqueous Barchans Collision Dynamical Equations).
To introduce the present skeleton model for 3D transverse dunes, we employ similar assumptions to those used in ABCDE;
\begin{figure}[htbp]
\begin{center}
\includegraphics[width=3.0 in]{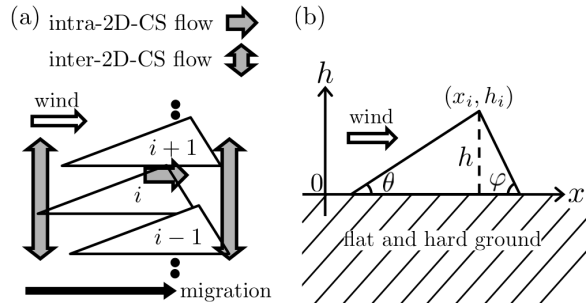}
\caption{
(a) Outlook of skeleton model;
triangular 2D-CSs $(1\le i \le N)$ constitute  a skeletonized 3D transverse dune or a 3D barchan.
(b) Shape similarity of 2D-CSs;
independent of their size.
Slope angles of $\theta$ and $\varphi$ are kept constant, then,
horizontal position and the height of $i$th 2D-CS are uniquely determined if  the coordinate of its crest $(x_i,h_i)$ is given.
}
\label{fig:sandflow}
\end{center}
\end{figure}

First, laterally arraying 2D wind directional cross-sections (hereafter 2D-CSs) of a 3D transverse dune (or a barchan) are set as elements of the present skeleton model.
Hence, the dynamics of each 2D-CS and the interaction between them is considered
(Fig. \ref{fig:sandflow}(a))

As mentioned above, 3D barchans are observed in dune fields with small amount of available sand, thus, they are isolated on a hard ground both in wind direction and lateral direction.
On the other hand, 3D transverse dunes observed under the same wind condition, extend in lateral direction therefore are not isolated.
However, in wind direction, successive two laterally extending crests of 3D transverse dunes are separated by inter-dunes hard ground  if the amount of available sand in 
the dune field is insufficient to cover all the ground,  and we treat such cases here. 
Shortly, we treat 2D-CSs constituting 3D barchans or 3D transverse dunes which are isolated in wind direction from windward and leeward  2D-CSs  
and each of them is assumed to have triangular shape located on hard and flat ground.
In addition, considering the fact that 2D-CSs of transverse dunes and barchans roughly have shape similarity independent of their size, we assume that the angles of their upwind and downwind slopes ($\theta$ and $\varphi$, respectively) are constant (Fig. \ref{fig:sandflow}(b)).
Then, geometrical constants A, B, C are introduced like,
\begin{eqnarray}
\label{eq:geoconst}
{\rm A}=\frac{\tan \theta \tan \varphi}{\tan \theta+\tan \varphi},{\rm B}=\frac{\tan \varphi}{\tan \theta+\tan \varphi},{\rm C}=\frac{\tan \theta}{\tan \theta+\tan \varphi},
\end{eqnarray}
where, the values of  ${\rm A,B}$ and ${\rm C}$ are set $\frac{1}{10}, \frac{4}{5}$ and $\frac{1}{5}$ respectively reflecting typical 2D-CS profiles of real barchans and transverse dunes. 
Based on above assumptions, the horizontal (i.e., wind directional) position and the height of each 2D-CS are uniquely determined if the coordinate $(x_i,h_i)$ $(1\le i\le N)$ of its crest is given(Fig. \ref{fig:sandflow}(b)).

\begin{figure}[t]
\begin{center}
\includegraphics[width=3.0 in]{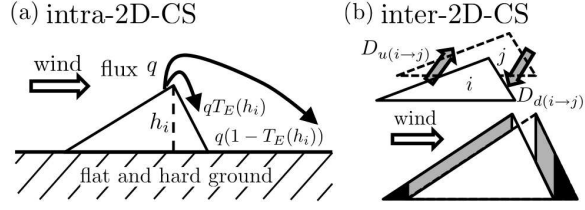}
\caption{
(a) Intra-2D-CS sand flow;
Over-crest sand flux $q$  and {\it sand trapping efficiency} $T_{E}(h_i)$ govern the intra-2D-CS sand flow. 
(b) Inter-2D-CS sand flow;
The dominant parts of the flow in the upwind and downwind slopes are proportional to 
the height difference of neighboring slopes multiplied by the overlapped length of these slopes 
as indicated by grey areas.
The additional contributions to the flow at the foot regions are indicated by black areas.
}
\label{fig:sandflux}
\end{center}
\end{figure}

Now, we consider the interaction between laterally neighboring 2D-CSs and then after the migration process of each.
Dune migration occurs by the sand flow along the surface.
In our description the sand flow is divided into two types:
(a) intra-2D-CS flow
(b) inter-2D-CS flow
(Fig. \ref{fig:sandflow}(a)).
In the intra-2D-CS flow, the most relevant quantity for the present modelling is the over-crest sand flux $q$ of each 2D-CS, which governs the erosion rate of its upwind slope. 
The over-crest sand flux $q$ also determines the deposition rate of the downwind slope, and it is noted that  finite ratio $1-T_{E}$ $(0\le 1-T_E \le 1)$ of $q$ is assumed to directly escape from the 2D-CS to the leeward inter-dune ground and the remaining ratio $T_{E}$ is deposited in the downwind slope. 
(Fig. \ref{fig:sandflux}(a)). This ratio $T_{E}$ is termed as the {\it sand trapping efficiency}, according to Momiji and Warren[10] .

The over-crest flux $q$ is assumed constant independent of the height of the 2D-CS,  while  $T_E$ is an increasing function of the height 
with limits $\displaystyle\lim_{h\to 0}T_{E}(h)=0,\lim_{h\to\infty}T_{E}(h)=1$,
with the specific form,
\begin{eqnarray*}
T_{E}(h)=h/(a+h),a=1.5.
\end{eqnarray*}

Next, the inter-2D-CS flux $D_{u(i\to j)}/D_{d(i\to j)}$ between upwind/downwind slopes of neighboring 2D-CSs, $i$ and $j$, is given as a function of heights and wind directional positions of their crests;

According to previous 3D dunes formation models\cite{kroy2002minimal}, 
we approximate that these inter-2D-CS fluxes are primarily contributed by the lateral  diffusion fluxes  integrated over upwind and downwind slopes respectively,  
thus, are proportional to the height difference between neighboring slopes multiplied by 
the overlapped length of  these slopes.   These quantities are shown as grey parallelograms in Fig. \ref{fig:sandflux}(b).

In addition,  we consider the contributions of the lateral diffusion at the foots of the upwind and downwind slopes  
where the diffusion flux is not proportional to the height difference between neighboring slopes but to 
the local height of the corresponding slopes because of no neighboring 2D-CS exists at these 
regions.  The diffused sand from these regions is assumed to be soon absorbed in the slopes of neighboring 2D-CS in short time. 
These quantities are shown as black triangles in Fig. \ref{fig:sandflux}(b).

Note that we treat,  in the below, only the cases of small height difference between neighboring slopes, therefore, 
the additional rule of diffusion at the foots-region of 2D-CSs does not affect the following results.

Specific forms of $D_{u(i\to j)}$ and $D_{d(i\to j)}$ are, 
\begin{subequations}
{\small
\begin{eqnarray}
\label{eq:Du}
D_{u(i\to j)}=\left\{
\begin{array}{ll}
\frac{\nu_u {\rm B}}{2{\rm A}}\{h_i^2 - [h_j-\frac{{\rm A}}{{\rm B}}(x_j-x_i)]^2\} & x_j-x_i>0\\
\frac{\nu_u {\rm B}}{2{\rm A}}\{[h_i+\frac{{\rm A}}{{\rm B}}(x_j-x_i)]^2 - h_j^2\} & x_j-x_i\le0\\
\end{array}
\right.\\
\label{eq:Dd}
D_{d(i\to j)}=\left\{
\begin{array}{ll}
\frac{\nu_d {\rm C}}{2{\rm A}}\{h_j^2 - [h_i-\frac{{\rm A}}{{\rm C}}(x_j-x_i)]^2\} & x_j-x_i>0\\
\frac{\nu_d {\rm C}}{2{\rm A}}\{[h_j+\frac{{\rm A}}{{\rm C}}(x_j-x_i)]^2 - h_i^2\} & x_j-x_i\le0.\\
\end{array}
\right.
\end{eqnarray}
}
\end{subequations}
where $\nu_u$ and $\nu_d$ are phenomenological parameters to control the amount of inter-2D-CS sand flow at respective sides of slopes.

Now we consider the migration of each 2D-CS.
As mentioned above, the size and the wind directional position of $i$th 2D-CS is uniquely determined if the coordinate $(h_i,x_i)$ of its crest is given.
Therefore to describe the dynamics of $(h_i,x_i)$ $(1\le i \le N)$ corresponds to give the skeletonized dynamics of a 3D dune.

\begin{figure}[b]
\begin{center}
\includegraphics[width=3.0 in]{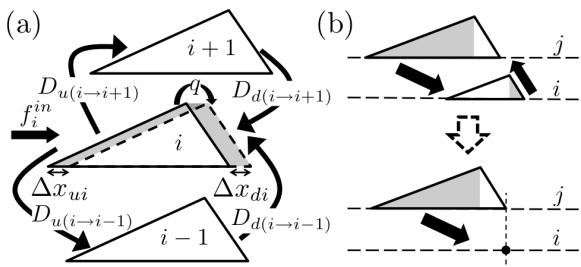}
\caption{
(a) Combinatorial action of intra-2D-CS flow $q$ and inter-2D-CS flow causes the erosion of upwind slope with width $\Delta x_{ui}$ and the deposition of downwind slope with the width $\Delta x_{di}$.
(b) Even after $i$th 2D-CS is vanished, its crest position $x_i$ is virtually kept allocated to define the inter-2D-CS flow from neighboring 2D-CS.
}
\label{fig:eq-vani}
\end{center}
\end{figure}

Here, $\Delta x_{ui}$ and $\Delta x_{di}$ are, respectively, the wind directional displacement of the upwind and the downwind slopes of $i$th 2D-CS during $\Delta t$
(Fig. \ref{fig:eq-vani}(a)), while the change of the coordinate $(h_i, x_i)$ of the 2D-CS crest within the same interval are denoted as $\Delta h_i$ and $\Delta x_i$, respectively.
Then, $\Delta h_i$ and $\Delta x_i$ are expressed by $\Delta x_{di}$ and $\Delta x_{ui}$,
\begin{subequations}
\begin{eqnarray}
\label{eq:101}
\Delta h_i = {\rm A}\Delta x_{di} - {\rm A}\Delta x_{ui}\\
\label{eq:102}
\Delta x_i = {\rm B}\Delta x_{di} + {\rm C}\Delta x_{ui}
\end{eqnarray}
\end{subequations}
where A, B, C are the geometrical constants introduced in (\ref{eq:geoconst}).
Consequently the eroded area per $\Delta t$ along the upwind slope on $i$th 2D-CS is expressed as,
{\small
\begin{eqnarray}
\label{eq:103}
\frac{2h_i+\Delta h_i}{2}\Delta x_{ui}=\Delta t (q+D_{u(i\to i-1)}+D_{u(i\to i+1)}+f_i^{in})
\end{eqnarray}
}where $q$ is the over-crest sand flux, $D_{u(i\to i-1)}$ and $D_{u(i\to i+1)}$ are inter-2D-CS flux from $i$th 2D-CS upwind slope to ($i-1$)th and ($i+1$)th 2D-CSs,respectively.
The quantity $f_i^{in}$ is incoming flux to $i$th 2D-CS from the windward inter-dune ground.
Similarly the deposited area per $\Delta t$ along the downwind slope on $i$th 2D-CS is expressed as,
{\small
\begin{eqnarray}
\label{eq:104}
\frac{2h_i+\Delta h_i}{2}\Delta x_{di}=\Delta t (qT_{E}(h_i)+D_{d(i\to i-1)}+D_{d(i\to i+1)})
\end{eqnarray}
}where $T_{E}(h_i)$ is the sand trapping efficiency, $D_{d(i\to i-1)}$ and $D_{d(i\to i+1)}$ are the 2D-CS flux from $i$th 2D-CS downwind slope to ($i-1$)th and ($i+1$)th 2D-CSs, respectively.
Using eqs. (\ref{eq:101})-(\ref{eq:104}) and taking their limits $\Delta t \to 0$ and $\Delta h_i \to 0$, a system of coupled ordinary equations,
\begin{subequations}
{\footnotesize
\begin{eqnarray}
\label{eq:dh}
\hspace{-6 ex}\frac{dh_i}{dt}=\frac{{\rm A}}{h_i}\left(q(T_{E}(h_i)-1)+\sum_{j=i\pm1}(D_{d(i\to j)}-D_{u(i\to j)})+f_i^{in}\right)\\
\label{eq:dx}
\hspace{-12 ex}\frac{dx_i}{dt}=\frac{1}{h_i}\left(q({\rm B}T_{E} (h_i)+{\rm C})+\sum_{j=i\pm1}({\rm B}D_{d(i\to j)}+{\rm C}D_{u(i\to j)})-f_i^{in}\right)\\
\nonumber
(1\le i \le N).
\end{eqnarray}
}
\end{subequations}
is obtained, which describes the dynamics of a skeletonized 3D dune.

Because dunes treated here are assumed to migrate on a flat and hard ground, its minimum height is zero that is not cared in (\ref{eq:dh}),(\ref{eq:dx}).
Therefore we add a rule for the vanishment of 2D-CSs, that is, if $h_i$ decreases to $h_i=0$, then $i$th 2D-CS is taken as vanished.
Note that the wind directional position $x$ of already vanished 2D-CS is virtually kept allocated at the foot of the downwind foot of the neighboring 2D-CS in order to determine the sand flux from the neighboring 2D-CSs according to eqs. (\ref{eq:Du}) and (\ref{eq:Dd})
(Fig. \ref{fig:eq-vani}(b)).

Numerical simulation of eqs. (\ref{eq:dh}),(\ref{eq:dx}) is conducted with $N=1000$ of 2D-CSs.
The lateral boundary condition is set periodic, whereas the boundary condition in wind direction is set more carefully.
Because the wind directional position  $x$ is  variable of  (\ref{eq:dx})  thus boundary condition for the 
2D-CSs is not required.
However, as mentioned above, finite ratio $1-T_E(h_i)$ of the over-crest flux $q$ escapes from each 2D-CS.
Then, to keep the realistic situation of dunes dynamics in desert fields where incoming sand flux to a dune is supplied by
escaping sand flux from windward dunes, the total amount of escaping sand,
\begin{eqnarray*}
F_{total} = \sum_{i=1}^N q(1-T_E(h_i)),
\end{eqnarray*}
is set uniformly redistributed to the upwind slopes of each 2D-CS.
It means that $f_i^{in}=F_{total}/N$ in (\ref{eq:dh}) and (\ref{eq:dx}).
Note that if finite number of 2D-CSs have already vanished they do not catch the redistributed sand, meaning that total amount of sand constituting 2D-CSs in the system is conserved only if all 2D-CSs are kept non-vanished, in that case, total area of 2D-CSs
\begin{eqnarray*}
S=\sum_{i=1}^N \frac{h_i^2}{2A}
\end{eqnarray*}
is kept constant too.

In the present model of eqs. (\ref{eq:dh}),(\ref{eq:dx}) accompanied with eqs. (\ref{eq:Du}),(\ref{eq:Dd}), three-environmental parameters $\nu_u,\nu_d,q$ are introduced all of which increase if the wind force increases.
Among them, $\nu_u,\nu_d$ are related to the increase of sand flow in the lateral direction, while $q$ is related to the sand flow in the wind direction.
Here ignoring the correlated increase of these three parameters responding to the increase of wind force, we independently vary them as individual control parameters.

As the initial condition of the numerical simulations, the height of 2D-CS crests are set uniform, i.e., $h_i(0)=H_0$$(0\le i \le N)$.
By varying this initial height we control the amount of available sand in the field.
Moreover, the initial wind directional position of crests are set wavy with small amplitude of sinuosity, that is, $x_i(0)=H_0/20 \sin(2\pi i/N)$.
We check if the initial amplitude of sinuosity in $x_i$ $(0\le i \le N)$ grows or not by measuring the quantity,
\begin{eqnarray*}
Var(t)=\sum_{i=1}^N(x_i(t)-x_{i+1}(t))^2 .
\end{eqnarray*}

\begin{figure}[b]
\begin{center}
\includegraphics[width=3.0 in]{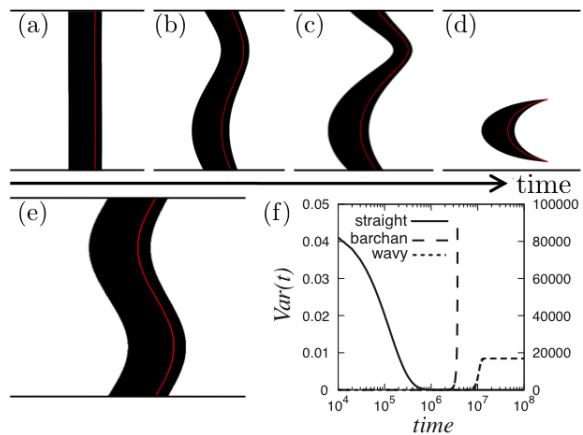}
\caption{
(a)$\sim$(d) Time evolutions from an initial straight transverse dune to a barchan.
(e) Stable state of wavy transverse dune.
(f) Typical time evolutions of the variance of dune crest line.
Depending on the values of control parameters, three phases are observed;
(I) ST-phase (real line): straight transverse dune is kept stable,
(II) WT-phase (short dashed line): wavy transverse dune is kept stable,
(III) B-phase (longer dashed line): initial transverse dune deforms into a barchan.
}
\label{fig:tra-bar}
\end{center}
\end{figure}

\begin{figure}[t]
\begin{center}
\includegraphics[width=3.5 in]{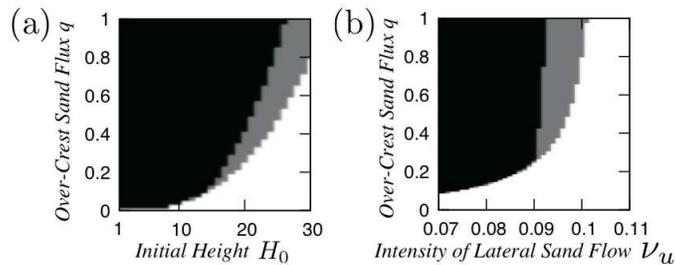}
\caption{
Results of the numerical simulations varying two pairs of control parameters (a) $H_0$ and $q$, (b) $\nu_u$ and $q$.
In white areas, ST-phase is realized, whereas, in grey areas WT-phase, in black areas B-phase is realized.
Other parameters are set like (a) $\nu_u=0.1,\nu_d=0.1$, (b) $\nu_d=0.1,H_0=30.0$.
}
\label{fig:simu}
\end{center}
\end{figure}

Simulations of the model exhibit three different phases depending on values of control parameter (Fig. \ref{fig:tra-bar}):
\begin{description}
\item[I)] If $Var(t)/Var(0)<10^{-2}$ is satisfied at $t=10^8$ we consider the laterally extending straight transverse dune is stable.
We call this 'ST-phase'.
\item[II)] If $Var(t)/Var(0)\ge 10^{-2}$ is satisfied at $t=10^8$ we take the wavy shape of a transverse dune as temporally stable.
We call this 'WT-phase'.
\item[III)] If, at least, one 2D-CS shrinks to become $h_i(t)=0$ within $t\le10^8$ thereafter the transverse dune will soon deform into the typical shape of barchan.
We call this 'B-phase'.
\end{description}
In the first simulation, we fix $\nu_u(=0.1)$ and $\nu_d(=0.1)$ and vary two parameters: $H_0$ and $q$
(Fig. \ref{fig:simu}(a)).
The increase of $H_0$ enhances the stability of the shape of straight transverse dune and the decrease of $H_0$ destabilizes its shape to enforce the deformation to a barchan.
This result qualitatively corresponds to the well known fact that barchans are formed in the field with small amount of available sand.

In the next simulation, we fix $\nu_d=0.1$ and set $H_0=30.0$ and vary two parameters, $\nu_u$ and $q$
(Fig. \ref{fig:simu}(b)).
As $\nu_u$ is set larger, the straight shape of transverse dune is more stabilized.
On the other hand the increase of $q$ destabilizes the transverse dune to enforce the deformation to barchan.
In this way the balance between the wind directional flow and the lateral flow determine the stability of the shapes of transverse dunes.

Finally, we conduct linear stability analysis of two 2D-CSs system which is considered as the simplest skeleton model of a transverse dune.
Here the uniformly redistributing rule of escaped sand flow is applied like the above simulations.
Thus, as long as two 2D-CSs are kept non-vanished, the total area of cross-sections $S$ is kept constant.
In this case,  (\ref{eq:dh}) and (\ref{eq:dx}) consist of four variables $(h_1,h_2,x_1,x_2)$, and $h_2$ is expressed like $h_2=\sqrt{2{\rm A}S-h_1^2}$.
Defining the wind directional relative crest position $y=x_2-x_1$ of two cross-sections, (\ref{eq:dh}) and (\ref{eq:dx}) are simplified into two variables coupled equations,
{\small
\begin{subequations}
\begin{eqnarray}
\label{eq:3}
\hspace{0 em}\frac{dh_1}{dt}=\frac{{\rm A}}{h_1}\left(q(T_E(h_1)-1)+D_{d(1\to 2)}-D_{u(1\to 2)}+f_1^{in}\right)\\
\label{eq:4}
\hspace{0 em}\frac{dy}{dt}=q{\rm B}\left(\frac{T_E(h_2)}{h_2}-\frac{T_E(h_1)}{h_1}\right)+({\rm C}-f_1^{in})\left(\frac{1}{h_2}-\frac{1}{h_1}\right)\nonumber \\
-({\rm B}D_{d(1\to 2)}-{\rm C}D_{u(1\to 2)})\left(\frac{1}{h_2}+\frac{1}{h_1}\right)
\end{eqnarray}
\end{subequations}
}
A fixed point $(h_1^*,y^*)=(\sqrt{{\rm A}S},0)$ of (\ref{eq:3}),(\ref{eq:4}) corresponds to the straightly extending state of a transverse dune.
We analyze the stability of this fixed point.
\begin{figure}[t]
\begin{center}
\includegraphics[width=3.5 in]{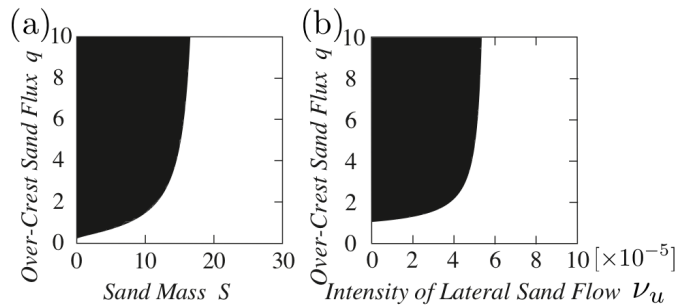}
\caption{
Results of the linear stability analysis of the fixed point of (\ref{eq:3}),(\ref{eq:4}) varying two pairs of control parameters: (a) $(S,q)$, (b) $(\nu_u,q)$.
White area indicate the conditions under which straightly extending transverse dune is stable, whereas in the black area such transverse dune is unstable.
Other parameter are set like (a) $\nu_u=5.0\times 10^{-5},\nu_d=1.0\times 10^{-4}$, (b) $\nu_d=1.0\times 10^{-4},S=15$.
}
\label{fig:cross2}
\end{center}
\end{figure}
Figure. \ref{fig:cross2}(a) and (b) show the linear stability of the fixed point in two sets of parameter spaces: ($S,q$) and ($\nu_u,q$) spaces.
In both figures, the fixed points are linearly stable in white areas, whereas they are unstable in black areas.
These results qualitatively corresponds to those shown in Fig. \ref{fig:simu}(a) and (b).

In this letter, we proposed a skeleton model of 3D dune dynamics and studied the stability of the shape of transverse dunes through numerical and analytical methods.
We also got a qualitative correspondence to the previous results obtained through observations of real dunes and more complicated simulation models.
Because of the simplicity of the model, we expect this model supplies us with an effective tool for the theoretical study for the complex dynamics of 3D dunes.
\vspace{1 ex}

\noindent
{\bf Acknowledgment}

This study was partially supported by the Global COE Program Formation and Development of Mathematicael Sciences Based on Modeling and Analysis.
We thank Dr Momiji in various comment.

\end{document}